\documentclass[12pt,showpacs,showkeys,amsmath,amssymb]{revtex4}
\usepackage{amsmath,amsfonts,amsthm,amscd,amssymb,latexsym}
\usepackage{bm}
\usepackage{dcolumn}
\usepackage{graphicx}
\usepackage{epstopdf}
\usepackage{color}
\usepackage{epsf}
\usepackage{epsfig}
\usepackage{graphicx, epic, eepic, color}

\begin{document}

\title{Rotational Doppler Effect and Spin-Rotation Coupling}

\author{Bahram \surname{Mashhoon}$^{1,2}$} 
\email{mashhoonb@missouri.edu}

\affiliation{
$^1$Department of Physics and Astronomy,
University of Missouri, Columbia,
Missouri 65211, USA \\
$^2$School of Astronomy,
Institute for Research in Fundamental Sciences (IPM),
Tehran 19395-5531, Iran
}

\date{\today}

\begin{abstract}
The Rotational Doppler Effect (RDE) involves both the orbital angular momentum of electromagnetic radiation as well as its helicity. The RDE phenomena associated with photon helicity go beyond the standard theory of relativity. The purpose of this paper is to elucidate the theoretical basis of the helicity-dependent RDE in terms of the general phenomenon of spin-rotation coupling. The physical implications of this coupling are briefly pointed out and the nonlocal theory of rotating observers is described. For an observer rotating uniformly about the direction of incidence of a plane wave of definite helicity, the nonlocally measured amplitude of the wave is larger (smaller) if the observer rotates in the same (opposite) sense as the electromagnetic field. The implications of the nonlocal helicity dependence of the measured amplitude of the radiation are briefly discussed. 
\end{abstract}


\maketitle

\section{Introduction}

The rotational Doppler effect (RDE) is a variant of the standard Doppler effect when the receiver rotates and is thus noninertial.  This aspect of the Doppler effect has now gained prominence in experimental physics; see, for instance,~\cite{LCT, LZZ, ZFD, FPW, DLLL, CWL, E+E} and the references cited therein. The theoretical basis of RDE has been widely discussed in the context of the quantum theory of the interaction of the electromagnetic radiation with matter; see, for instance,~\cite{B+B, BSV, LBS, SLP, ABM} and references therein. The main purpose of the present work is to discuss RDE in connection with the relativistic physics of rotating observers. Accelerated systems play a fundamental role in the interface connecting the physics of the special and general theories of relativity. It is therefore necessary to discuss carefully the extension of Lorentz invariance to accelerated observers.  In the rest of this introductory section, the derivation of the relativistic Doppler effect based on Lorentz invariance is presented. The standard extension of Lorentz invariance to accelerated systems in Minkowski spacetime is discussed in Section II. The limitations of the standard theory are pointed out and attempts to mitigate them  are discussed in Sections III-V. Section VI contains a brief discussion of our results. 

In a global inertial frame in Minkowski spacetime with Cartesian coordinates $x^\mu = (ct, \mathbf{x})$, we consider the propagation of an electromagnetic wave packet.  Inertial observers that are spatially at rest in this background inertial frame can determine via Fourier analysis that the packet consists of a spectrum of plane electromagnetic waves each with null propagation vector $k^\mu = (\omega/c, \mathbf{k})$, where $\omega = c |\mathbf{k}|$. We are interested in the reception of the radiation by the inertial observers that are all at rest in a global inertial frame moving with unit 4-velocity vector $u^\mu = \gamma (1, \mathbf{v}/c)$. Here, $u^\mu = dx^\mu /d\tau$, where $d\tau$ is given by 
\begin{equation}\label{I1}
d\tau = c\,\left(1 - \frac{v^2}{c^2}\right)^{1/2} dt\,,
\end{equation}
 $v = |\mathbf{v}|$  and  $\gamma  = (1 - v^2/c^2)^{-1/2}$ is the Lorentz factor. The Lorentz transformation from the background frame to the moving frame with Cartesian coordinates $x'^\mu = (ct', \mathbf{x}')$ that is a pure boost with no rotation is given by
\begin{equation}\label{I2}
t = \gamma \left(t' + \frac{1}{c^2}\mathbf{v}\cdot \mathbf{x}'\right)\,,
\end{equation}
\begin{equation}\label{I3} 
\mathbf{x} = \mathbf{x}'+\frac{\gamma - 1}{v^2}(\mathbf{v}\cdot \mathbf{x}')\mathbf{v} - \frac{1}{c^2} \gamma \mathbf{v}t'\,.
\end{equation} 
Here, we have assumed that at $t = t' = 0$, the spatial origins of the two systems coincide, namely, $\mathbf{x} = \mathbf{x}' = 0$.  We define the invariant proper time $\tau$ of an inertial observer to be the time it measures in its own rest frame multiplied by the speed of light $c$. For instance, it follows from Eq.~\eqref{I2} that $c dt' = d\tau = c dt/\gamma$ for the moving observer fixed at $\mathbf{x}'$, in agreement with Eq.~\eqref{I1}.  
 
The inertial observers at rest in the moving frame measure a similar wave spectrum with components $k'^\mu = (\omega'/c, \mathbf{k}')$. The Lorentz transformation connecting the measurements of the two classes of inertial observers, each class at rest in a global inertial frame, leaves the phase $\phi$,
\begin{equation}\label{I4}
\phi (t, \mathbf{x}) = - \omega t + \mathbf{k}\cdot \mathbf{x}\,,
\end{equation}
of each plane-wave component of the packet invariant; that is, $\phi (t, \mathbf{x}) = \phi' (t', \mathbf{x}')$ or 
 \begin{equation}\label{I5}
-\omega t + \mathbf{k}\cdot \mathbf{x} = -\omega' t' + \mathbf{k}'\cdot \mathbf{x}'\,.
\end{equation}
From this phase invariance and Eqs.~\eqref{I2} and~\eqref{I3}, one can simply derive the standard relativistic formulas for the Doppler effect and aberration of electromagnetic radiation, namely, 
\begin{equation}\label{I6}
\omega' = \gamma (\omega - \mathbf{v}\cdot \mathbf{k})\,,
\end{equation}
\begin{equation}\label{I7}
\mathbf{k}' = \mathbf{k}+\frac{\gamma - 1}{v^2}(\mathbf{v}\cdot \mathbf{k})\mathbf{v} - \frac{1}{c^2} \gamma \omega \mathbf{v}\,.
\end{equation}
In particular, we note that the Doppler effect has to do with the change in the frequency of the wave due to the relative motion of the source and the receiver, while aberration involves the change in the measured direction of propagation of the wave.

Let us note that we can write the Doppler formula as
\begin{equation}\label{I8}
\omega' = - k_{\mu}u^\mu\,,
\end{equation}
which illustrates the fact that what is measured is an invariant under Lorentz transformations. 

We now turn to the standard theory of noninertial systems.

\section{Measurements of Accelerated Observers} 

To simplify matters, we generally assume in this paper that the gravitational interaction is turned off; therefore, the spacetime under consideration is Minkowskian. In general, an observer in Minkowski spacetime has access to an orthonormal tetrad $e^{\mu}{}_{\hat \alpha}$,  such that $e^{\mu}{}_{\hat 0} = dx^\mu / d\tau$ is the 4-velocity of the observer and $\tau$ is its proper time; moreover, $e^{\mu}{}_{\hat i}$, $i = 1, 2, 3$, constitute the observer's local spatial frame. Hatted indices enumerate the tetrad axes in the local tangent space. The orthonormality condition can be expressed as
\begin{equation}\label{J1}
\eta_{\mu \nu} \,e^{\mu}{}_{\hat \alpha}\,e^{\mu}{}_{\hat \beta} = \eta_{\hat \alpha \hat \beta}\,,
\end{equation}
where $\eta_{\mu \nu}$ is the Minkowski metric tensor and is given by diag$(-1, 1, 1, 1)$ in our convention. 

For the hypothetical ideal inertial observers, we have
\begin{equation}\label{J2}
\frac{de^{\mu}{}_{\hat \alpha}}{d\tau} = 0\,.
\end{equation}
The temporal variation of a component of an observer's tetrad frame would indicate that the observer is noninertial and is therefore subject to translational or rotational acceleration. Actual observers in Minkowski spacetime are all accelerated. At each instant of proper time $\tau$, the tetrad frame of an accelerated observer $e^{\mu}{}_{\hat \alpha}(\tau)$ also belongs to an otherwise identical momentarily comoving \emph{inertial} observer and it is natural to assume that their measurements of physical quantities give the same results at that instant. In this way, Lorentz invariance has been extended to accelerated systems in the standard theory of special relativity~\cite{Ein}. We discuss this approach in some detail in the rest of this section. 

\subsection{Locality}

The standard theory of relativity is based on two basic assumptions: Lorentz invariance and the hypothesis of locality. Lorentz invariance is a basic symmetry of nature and connects the measurements of hypothetical ideal inertial observers in Minkowski spacetime. Inertial observers remain forever at rest in global inertial frames of reference and have been of fundamental significance for the formulation of the basic laws of physics since their introduction by Isaac Newton. However, all real observers are more or less accelerated. Therefore, we need to discuss the measurements of accelerated observers in Minkowski spacetime. Let us recall that an accelerated Newtonian point particle is always momentarily inertial; that is, the classical particle follows the law of inertia if its acceleration is instantaneously turned off. Hence, an accelerated path is the envelope of the osculating straight lines tangent to the trajectory. The Newtonian theory of accelerated observers can be straightforwardly extended to the relativistic regime by replacing Galilean invariance with Lorentz invariance. The accelerated observer is thus momentarily equivalent to an otherwise identical instantaneously comoving inertial observer. Therefore, Lorentz transformation can be applied point by point along the world line of the accelerated observer in order to determine what the accelerated observer measures.  In other words, the accelerated observer has a velocity $\mathbf{v}(t)$ at each instant of time $t$ and the instantaneous Lorentz transformation associated with 
$\mathbf{v}(t)$ can be employed to connect the accelerated observer to inertial observers; in this way, relativity can be extended to accelerated systems~\cite{Ein}. 

In this ``small-scale" approach~\cite{ROB}, the acceleration of the observer is ignored locally, but is taken into consideration via the variation in the velocity of motion embodied in the pointwise Lorentz transformations. This hypothesis of locality determines what accelerated observers measure according to the standard special theory of relativity~\cite{Ein}. Measuring devices that work in accordance with this basic assumption will be called ``standard". 
 Measuring devices carried by accelerated observers are subject to various inertial forces and effects that could possibly integrate over proper time to render such devices non-standard. Nevertheless, we limit our considerations throughout to measuring devices that are robust enough to function as standard instruments to a sufficient degree of accuracy.  In connection with the notion of standard rods and clocks, it is interesting to note that Einstein made a similar assumption in the footnote on page 60 of his book~\cite{Einstein}.

The locality postulate implies that the accelerated observer along its world line passes continuously through an infinite sequence of inertial observers and their associated inertial frames. The Lorentz transformation connecting one such inertial \emph{rest} frame with Cartesian coordinates $x'^\mu = (ct', x', y', z')$  to the background inertial frame with Cartesian coordinates $x^\mu = (ct, x, y, z)$ leaves the Minkowski spacetime interval,
\begin{equation}\label{J3}
ds^2 = \eta_{\mu \nu} dx^\mu dx^\nu\,,
\end{equation}
invariant. Therefore, 
\begin{equation}\label{J4}
c^2 dt^2 - dx^2 - dy^2 - dz^2  =  c^2 dt'^2\,,
\end{equation}
where $d\mathbf{x}' = 0$ and $dt'$ is the infinitesimal time interval experienced by the accelerated observer  in the comoving inertial frame at time $t'$  in accordance with the locality postulate. From $d\tau = c \,dt'$, we find the proper time measured by the \emph{standard clock} carried by the accelerated observer, namely,
\begin{equation}\label{J5}
\tau = c \int_0^t \left[1 - \frac{v^2(\xi)}{c^2}\right]^{1/2} d\xi\,,
\end{equation}
where $\mathbf{v}(t)$ is the velocity of the observer with respect to the background global inertial frame and we have assumed that $\tau = 0$ when $t = 0$. That is, the accelerated observer's proper time $\tau$ is the accumulation of local infinitesimal comoving time intervals. If the accelerated observer travels from an initial event $x_{I}^\mu = (c t_I, \mathbf{x})$ in the background inertial frame and finally returns at a later time to the same place, i.e. $x_{F}^\mu = (c t_F, \mathbf{x})$, we find
\begin{equation}\label{J6}
\frac{1}{c} (\tau_F - \tau_I) = \int_{t_I}^{t_F} \left[1 - \frac{v^2(t)}{c^2}\right]^{1/2} dt < t_F - t_I\,,
\end{equation}
where $t_F - t_I$ is the elapsed temporal interval according to an inertial observer at rest in the background global inertial frame in Minkowski spacetime. In this way, Einstein explained away the twin paradox~\cite{Ein}. 

Next, imagine the reception of electromagnetic radiation by the accelerated observer. The instantaneous Lorentz transformation connecting the background global frame with the local inertial rest frame of the accelerated observer affects the field amplitudes, but the phase of the wave remains invariant, i.e. $d\phi = \eta_{\mu \nu} k^\mu dx^\nu$, where $k^\mu = (\omega/c, \mathbf{k})$ is the propagation 4-vector of the wave in the background frame. That is, $d\phi = \eta_{\mu \nu} k'^\mu dx'^\nu$, where
\begin{equation}\label{J7}
\omega = - \left(\frac{\partial \phi}{\partial t}\right)_{\mathbf{x}}\,, \qquad \mathbf{k} =  \left(\frac{\partial \phi}{\partial \mathbf{x}}\right)_{t}\,
\end{equation}
and
\begin{equation}\label{J7a}
\omega' = - \left(\frac{\partial \phi}{\partial t'}\right)_{\mathbf{x}'}\,, \qquad \mathbf{k}' =  \left(\frac{\partial \phi}{\partial \mathbf{x}'}\right)_{t'}\,.
\end{equation}

The relativistic Doppler effect and aberration for accelerated observers follow from phase invariance. In particular, the corresponding Doppler frequency is given by
\begin{equation}\label{J8}
\omega'_{\rm D}(\tau)  = \gamma [\omega - \mathbf{v}(t) \cdot \mathbf{k}]\,,
\end{equation}
where $\mathbf{v}(t)$ is the instantaneous velocity of the accelerated observer. The locally measured frequency~\eqref{J8} can also be expressed in the form of an invariant as in Eq.~\eqref{I8}; that is, $\omega'_{\rm D}(\tau)$ is proportional to the projection of the incident propagation 4-vector of the wave upon the 4-velocity vector of the accelerated observer. 

The basic limitations of the locality postulate have been studied in detail~\cite{BM1, BM2, Mashhoon:2003st}; indeed, this postulate is physically reasonable if the motion is sufficiently slow such that the velocity of the accelerated system is essentially constant during an elementary act of measurement. In connection with the approximation involved in the use of standard rods and clocks in the measurement process, there is a brief discussion in Weyl's book~\cite{Weyl}. After pointing out that  under acceleration the classic rigid bodies no longer exist in relativity theory due to the finite speed of interactions, Weyl discussed accelerated observers on pages 176 and 177 of his book~\cite{Weyl} and concluded that the approximation basically involved slow steady change in the observer's velocity. 

A curved path is the envelope of the straight lines tangent to the path; similarly, a pointlike accelerated observer passes through an infinite sequence of otherwise identical momentarily comoving inertial observers. Is this geometric fact also true for a physical observer that carries along measuring instruments? Standard ideal rods and clocks are assumed to work in accordance with the hypothesis of locality. More generally, we assume throughout that accelerated observers have access to \emph{ideal standard measuring devices}.  This approach is consistent with the notion that measuring devices generally follow the laws of classical physics~\cite{Mashhoon:2008vr}.

The temporal dependence of the measured frequency $\omega'_{\rm D}(\tau)$ in Eq.~\eqref{J8} indicates that the domain of applicability of the Doppler formula for accelerated observers is rather limited~\cite{Mashhoon:2008vr, BMB}. Rotating observers will be considered in the rest of this paper. 

\subsection{Doppler Effect for Rotating Observers}

Let us consider an observer that rotates about the $z$ axis in a global inertial frame in Minkowski spacetime. The observer is located at position $\mathbf{r}$ and rotates with angular velocity $\boldsymbol{\Omega}= \Omega \,\hat{\mathbf{z}}$ and velocity $\mathbf{v} = \boldsymbol{\Omega} \times \mathbf{r}$, where $\hat{\mathbf{z}}$ is a unit vector along the $z$ axis. 

The application of the relativistic Doppler formula~\eqref{J8} to the motion of the rotating observer results in  
\begin{equation}\label{J9}
\omega_{\rm D}' = \gamma [\omega - (\boldsymbol{\Omega} \times \mathbf{r}) \cdot \mathbf{k}] = \gamma (\omega - \boldsymbol{\Omega} \cdot \boldsymbol{\ell})\,,
\end{equation}
where $\boldsymbol{\ell} = \mathbf{r} \times \mathbf{k}$. Multiplying Eq.~\eqref{J9} by $\hbar$, we get 
\begin{equation}\label{J10}
E' =  \gamma (E - \boldsymbol{\Omega} \cdot \mathbf{L})\,,
\end{equation}
where 
\begin{equation}\label{J11}
E =  \hbar\, \omega\,, \qquad  \mathbf{L} = \hbar\, \boldsymbol{\ell}\,.
\end{equation}
Here, $E$ is the energy and $\mathbf{L}$ is the orbital angular momentum of the light ray with wave vector $\mathbf{k}$. Equation~\eqref{J10} is the relativistic generalization of the well-known classical mechanical result for the Hamiltonian in a rotating frame of reference~\cite{L+L, Mashhoon:2021qtc}. It follows that the orbital angular momentum aspects of the RDE are in accord with the special theory of relativity.

On the other hand, the generator of rotations is the total angular momentum $\mathbf{J} = \mathbf{L}  + \mathbf{S}$, where $\mathbf{S} = \pm \hbar\, \hat{\mathbf{k}}$ represents the photon spin. Therefore, we expect
\begin{equation}\label{J12}
E' =  \gamma (E - \boldsymbol{\Omega} \cdot \mathbf{L} - \boldsymbol{\Omega} \cdot \mathbf{S})\,,
\end{equation}
which is the more general result giving rise to the helicity-rotation coupling that is part of the observational basis of the rotational Doppler effect (RDE), but which goes beyond the standard relativistic Doppler effect. The next section is devoted to a detailed discussion of the photon helicity-rotation coupling. 

\section{Beyond Locality: Helicity-Rotation Coupling}

The main purpose of this paper is to point out that the helicity-rotation aspect of RDE goes beyond the locality postulate and phase invariance. The standard special theory of relativity is based on the hypothesis of locality;  therefore, it is necessary to provide a proper theoretical basis for the observed helicity-rotation coupling. We show that the validity of the locality hypothesis is limited to the WKB approximation. Locality is clearly valid for pointlike coincidences of classical point particles and rays of radiation. According to the Huygens principle, wave phenomena are intrinsically nonlocal; indeed,  this nonlocal wave aspect of RDE is illustrated in this section.

\subsection{Frequency Measurements of a Noninertial Observer}

Imagine a global inertial frame with Cartesian coordinates $x^\mu = (ct, x, y, z)$ in Minkowski spacetime and an observer that is at rest at the origin of spatial coordinates. The observer is noninertial, since it refers its measurements to axes that rotate with constant angular speed $\Omega > 0$ about the $z$ axis. The rotating axes therefore have unit directions 
\begin{equation}\label{L1}
\hat{\mathbf{x}}' = \hat{\mathbf{x}}\,\cos \Omega t + \hat{\mathbf{y}}\,\sin \Omega t\,, \quad \hat{\mathbf{y}}' = -  \hat{\mathbf{x}}\,\sin \Omega t +  \hat{\mathbf{y}}\, \cos \Omega t\,, \quad  \hat{\mathbf{z}}' = \hat{\mathbf{z}}\,
\end{equation}
and the observer employs coordinates $x'^\mu = (\tau, x', y', z')$, where $\tau = c t$ is the proper time of the noninertial observer and 
\begin{equation}\label{L2}
x' = x\, \cos \Omega t +  y\,\sin \Omega t\,, \quad y' = -  x\,\sin \Omega t +  y\,\cos \Omega t\,, \quad  z' = z\,.
\end{equation}

Consider now a plane electromagnetic wave of frequency $\omega$ and wave vector $\mathbf{k} = (\omega/c) \hat{\mathbf{z}}$ incident along the $z$ direction. The electromagnetic 4-potential $A_\mu$ of the wave is given by 
\begin{equation}\label{L3}
A_0 = 0\,, \quad \mathbf{A} = a_{\pm} (\hat{\mathbf{x}} \pm i \hat{\mathbf{y}}) e^{-i\,\omega (t - z/c)}\,,
\end{equation}
where $a_{\pm}$ is the complex constant wave amplitude and the upper (lower) sign indicates incident positive (negative) helicity radiation. Only linear fields are considered throughout; therefore, we use complex fields and adhere to the convention that their    real parts are physically significant. Throughout this paper, we employ for the sake of simplicity the vector potential $A_\mu$ instead of the electromagnetic field tensor $F_{\mu \nu}$, since it turns out that they give the same results regarding the frequency content of the field. Moreover, we work in the radiation gauge, where $A^\mu = (0, \mathbf{A})$ and $\nabla \cdot \mathbf{A} = 0$. 

Equation~\eqref{L3} represents the vector potential according to the fundamental ideal inertial observers at rest in the background global inertial frame; that is, what is measured is the projection of the vector potential upon the tetrad frame of the inertial observer, $e^{\mu}{}_{\hat \alpha}$, 
\begin{equation}\label{L4}
A_{\hat \alpha} = A_\mu\, e^{\mu}{}_{\hat \alpha}\,, \qquad   e^{\mu}{}_{\hat \alpha} = \delta^\mu_\alpha\,.
\end{equation}

However, for the measurements of the noninertial observer, 
\begin{equation}\label{L5}
A'_{\hat \alpha} = A_\mu\, e'^{\mu}{}_{\hat \alpha}\,;
\end{equation}
that is, the vector potential $A_\mu$ is projected upon the orthonormal tetrad frame $e'^{\mu}{}_{\hat \alpha}$ adapted to the noninertial observer. The corresponding tetrad axes are given explicitly in $x^\mu = (ct, x, y, z)$ coordinates by
\begin{equation}\label{L6}
 e'^{\mu}{}_{\hat 0} = (1, 0, 0, 0)\,,
\end{equation}
\begin{equation}\label{L7}
 e'^{\mu}{}_{\hat 1} = (0, \cos\Omega t, \sin\Omega t, 0)\,,
\end{equation}
\begin{equation}\label{L8}
 e'^{\mu}{}_{\hat 2} = (0, -\sin\Omega t, \cos\Omega t, 0)\,,
\end{equation}
\begin{equation}\label{L9}
 e'^{\mu}{}_{\hat 3} = (0, 0, 0, 1)\,.
\end{equation}
The adapted tetrad frame is thus the spacetime generalization of the uniformly rotating unit triad~\eqref{L1} employed by the noninertial observer. Using Eqs.~\eqref{L5}--\eqref{L9}, we find $A'_{\hat 0} = 0$, $A'_{\hat 3} = A_3$ and 
\begin{equation}\label{L10}
 A'_{\hat 1} = A_1 \cos\Omega t+A_2 \sin\Omega t\,,
\end{equation}
\begin{equation}\label{L11}
 A'_{\hat 2} =  - A_1 \sin\Omega t+A_2 \cos\Omega t\,.
\end{equation}
Next, from Eq.~\eqref{L3} we find
\begin{equation}\label{L12}
A'_{\hat 1} = a_{\pm}\, e^{-i(\omega \mp \Omega)t}\,,
\end{equation}
\begin{equation}\label{L13}
A'_{\hat 2} = \pm i\,a_{\pm}\, e^{-i(\omega \mp \Omega)t}\,,
\end{equation}
\begin{equation}\label{L14}
A'_{\hat 3} = 0\,.
\end{equation}

The noninertial observer refers its observations to axes rotating with \emph{constant} angular velocity $\Omega$ about the $z$ axis. Consequently, the local spacetime of the noninertial observer is \emph{stationary}; therefore, Fourier analysis of the measured vector potential reveals that 
\begin{equation}\label{L15}
\omega' = \omega \mp \Omega\,,
\end{equation}
where the upper (lower) sign refers to positive (negative) helicity radiation.  In the  positive (negative) helicity radiation that propagates along the $z$ direction, the electric and magnetic fields rotate in  the positive (negative) sense about the direction of propagation with angular speed $\omega$. These fields are then naturally perceived by the noninertial observer to be rotating with angular speed $\omega - \Omega$ ($\omega + \Omega$) in the positive (negative) sense about the $z$ axis. This composition of angular speeds is reminiscent of the composition of linear speeds in the Doppler effect. 

Taking the intuitive explanation of Eq.~\eqref{L15} one step further, it becomes clear that for an incident linearly polarized wave, the plane of linear polarization would appear to rotate in the negative sense about the $z$ axis with angular speed $\Omega$ as perceived by the noninertial observer.  This circumstance is the analogue of the magneto-optic Faraday effect via the Larmor theorem~\cite{Larmor}.  

For the noninertial observer under consideration here, its world line is the temporal axis and the hypothesis of locality implies that the measured Doppler frequency is
\begin{equation}\label{L16}
\omega'_{\rm D} = \omega\,,
\end{equation}
since the noninertial observer is fixed at $(x, y, z) = 0$ and $\mathbf{v} = 0$ in this case. Writing 
\begin{equation}\label{L17}
\omega' = \omega'_{\rm D} \left(1 \mp \frac{\Omega}{\omega}\right)\,, \qquad   \frac{\Omega}{\omega} = \frac{\lambdabar}{c/\Omega}\,,
\end{equation}
where $\lambdabar$ is the reduced wavelength of the incident radiation, we recognize that the helicity-dependent part of the RDE is a wave effect and hence goes beyond the ray (WKB) picture of the standard relativistic RDE. Moreover, multiplication of Eq.~\eqref{L15} by $\hbar$ results in 
\begin{equation}\label{L18}
E' =  E \mp \hbar\,\Omega = E - \mathbf{S} \cdot \boldsymbol{\Omega}\,,
\end{equation}
which illustrates the helicity-rotation coupling that is responsible for the deviation from the locality postulate.

\subsection{Oblique Incidence}

Let us now consider the case of oblique incidence. With no loss in generality, we can assume an incident plane electromagnetic wave of frequency $\omega$ and unit wave vector 
\begin{equation}\label{L19}
 \hat{\mathbf{k}} = (-\sin \theta, 0, \cos \theta)\,
\end{equation}
in Cartesian $(x, y, z)$ coordinates, where $\theta = 0$ corresponds to the previous case of incidence along the $z$ axis. The wave has definite helicity and electromagnetic 4-potential $A_\mu$ given by 
\begin{equation}\label{L20}
A_0 = 0\,, \quad \mathbf{A} = a_{\pm} (\hat{\mathbf{n}} \pm i \hat{\mathbf{y}}) e^{-i\,(\omega/c) (ct + \sin \theta \,x -  \cos \theta \,z)}\,,
\end{equation}
where the unit vector
\begin{equation}\label{L21}
 \hat{\mathbf{n}} = (\cos \theta, 0, \sin \theta)\,
\end{equation}
is an axis in the orthonormal triad $( \hat{\mathbf{n}}, \hat{\mathbf{y}},  \hat{\mathbf{k}})$. 

For the measurements of the noninertial observer fixed at the origin of spatial coordinates, we find $A'_{\hat 0} = 0$ and, instead of Eqs.~\eqref{L12}--\eqref{L14}, 
\begin{equation}\label{L22}
A'_{\hat 1} = a_{\pm}\, e^{-i\,\omega t}(\cos \theta \cos \Omega t \pm i\, \sin \Omega t)\,,
\end{equation}
\begin{equation}\label{L23}
A'_{\hat 2} = a_{\pm}\, e^{-i\,\omega t}(-\cos \theta \sin \Omega t \pm i\, \cos \Omega t)\,,
\end{equation}
\begin{equation}\label{L24}
A'_{\hat 3} = a_{\pm}\, e^{-i\,\omega t}\,\sin \theta\,.
\end{equation}
Fourier analysis of these fields reveals the presence of three frequencies 
\begin{equation}\label{L25}
\omega'_{m'} = \omega - m'\, \Omega\,, \qquad m' = +1, -1, 0\,;
\end{equation}
moreover, Eqs.~\eqref{L22}--\eqref{L24} can be expressed in terms of amplitudes $\psi_{(\hat i, m')}$ as
\begin{equation}\label{L22a}
A'_{\hat 1} = \psi_{(\hat 1, +1)}\, e^{-i\,\omega'_{+1} t} + \psi_{(\hat 1, -1)}\, e^{-i\,\omega'_{-1} t}\,,
\end{equation}
\begin{equation}\label{L23a}
A'_{\hat 2} = \psi_{(\hat 2, +1)}\, e^{-i\,\omega'_{+1} t} + \psi_{(\hat 2, -1)}\, e^{-i\,\omega'_{-1} t}\,
\end{equation}
and
\begin{equation}\label{L24a}
A'_{\hat 3} = \psi_{(\hat 3, 0)}\, e^{-i\,\omega'_{0} t}\,.
\end{equation}
That is, the component of the 4-potential along $e'^{\mu}{}_{\hat 1}$ contains frequencies $\omega'_{+1} = \omega - \Omega$ and $\omega'_{-1} = \omega + \Omega$ with amplitudes
\begin{equation}\label{L26}
\psi_{(\hat 1, +1)} = \pm a_{\pm} \frac{1\pm \cos \theta}{2}\,, \qquad \psi_{(\hat 1, -1)} =  \mp a_{\pm} \frac{1\mp \cos \theta}{2}\,,
\end{equation}
respectively, while along $e'^{\mu}{}_{\hat 2}$ we have the same frequencies but with amplitudes 
\begin{equation}\label{L27}
\psi_{(\hat 2, +1)} = \pm i\,a_{\pm} \frac{1\pm \cos \theta}{2}\,, \qquad \psi_{(\hat 2, -1)} =  \pm i\, a_{\pm} \frac{1\mp \cos \theta}{2}\,,
\end{equation}
respectively. Finally, along $e'^{\mu}{}_{\hat 3}$, the observer finds frequency $\omega'_{0} = \omega$ with amplitude
\begin{equation}\label{L28}
\psi_{(\hat 3, 0)} = a_{\pm} \sin \theta\,. 
\end{equation}

The average frequency measured by the noninertial observer may be calculated in the manner of wave mechanics as
\begin{equation}\label{L29}
< \omega' > \, = \frac{\sum \omega'_{m'} |\psi_{(\hat i, m')}|^2}{\sum |\psi_{(\hat i, m')}|^2}\,, 
\end{equation}
where the summation is over $i = 1, 2, 3$ and $m' = -1, 0, 1$. A simple calculation reveals that the numerator in Eq.~\eqref{L29} is given by $2 |a_{\pm}|^2 ( \omega \mp \cos \theta \,\Omega)$ and the corresponding denominator is given by $2 |a_{\pm}|^2$.
Hence, we find
\begin{equation}\label{L30}
< \omega' > \, = \omega \mp \cos \theta \,\Omega\,. 
\end{equation}
Multiplication by $\hbar$ again reveals the coupling of helicity with rotation
\begin{equation}\label{L31}
< E' > \, = E - \mathbf{S} \cdot \boldsymbol{\Omega}\,,
\end{equation}
where $\mathbf{S} = \pm \hbar \, \hat{\mathbf{k}}$ for the photon. The same result can be obtained from a detailed WKB treatment~\cite{Mashhoon:2002fq, Hauck:2003gy}.  

Let us note that we can write Eq.~\eqref{L29} in the form
\begin{equation}\label{L32}
< \omega' > \, = (\omega - \Omega)P_{+} + \omega P_{0} + (\omega + \Omega) P_{-}\,, 
\end{equation}
where the probability that the observer measures $\omega'_{m'}$ for $m' = +1, 0,  -1$, is given by $P_{+}$, $P_{0}$ and $P_{-}$, respectively. Here, 
\begin{equation}\label{L33}
P_{+} =  \frac{1}{4}(1\pm \cos \theta)^2\,, \qquad P_{-} = \frac{1}{4} (1\mp \cos \theta)^2\,, \qquad P_{0} = \frac{1}{2} \sin^2 \theta\,,
\end{equation}
 where $P_{+} + P_{-} + P_0 = 1$. In these relations, the upper (lower) sign corresponds to positive (negative) helicity radiation that is incident along the $\hat{\mathbf{k}}$ direction given by Eq.~\eqref{L19}. Let us recall that the eigenstates of the incident particle of spin $\hbar$ with respect to the $(\hat{\mathbf{n}}, \hat{\mathbf{y}},  \hat{\mathbf{k}})$ system can be transformed to the $(\hat{\mathbf{x}}, \hat{\mathbf{y}},  \hat{\mathbf{z}})$ system of the observer via the well-known matrix $(\mathcal{D}^j_{m m'})$ for $j= 1$ that is given by
\begin{equation}\label{L34}
 \left[
\begin{array} {ccc}
\frac{1}{2} (1+\cos \theta) & -\frac{1}{\sqrt{2}} \sin \theta & \frac{1}{2} (1-\cos \theta) \cr
\frac{1}{\sqrt{2}} \sin \theta  & \cos \theta & -\frac{1}{\sqrt{2}} \sin \theta \cr
\frac{1}{2} (1-\cos \theta) & \frac{1}{\sqrt{2}} \sin \theta & \frac{1}{2} (1+\cos \theta) \cr
\end{array}
 \right]\,;
\end{equation}
see, e.g.,  page 57 of  Ref.~\cite{Edmonds}. The application of this matrix on the helicity states of the incident photon implies that the amplitude for the spin states along the $z$ axis for an incident photon of definite helicity is proportional to
\begin{equation}\label{L35}
 \left[
\begin{array} {c}
\frac{1}{2} (1\pm\cos \theta)  \cr
\pm\,\frac{1}{\sqrt{2}} \sin \theta   \cr
\frac{1}{2} (1\mp\cos \theta)  \cr
\end{array}
 \right]\,
\end{equation}
based on the first and third columns of matrix~\eqref{L34}. We emphasize that the upper (lower) sign here indicates \emph{incident} positive (negative) helicity waves along the $\hat{\mathbf{k}}$ axis. The squares of the first and third elements of~\eqref{L35}, namely, $P_{+}$ and $P_{-}$,  represent the probability that the photon spin is $+\hbar$ and $-\hbar$ along the $z$ axis, i.e. the rotation axis of the observer, respectively, while $P_{0}$, which is equal to the square of the second element of~\eqref{L35}, is the probability that the photon spin along the $z$ axis vanishes.  In this way, Eq.~\eqref{L32} can provide an intuitive understanding of why the average frequency measured by the noninertial observer depends upon the helicity of the incident radiation.  

\subsection{Rotating Observers}

Consider an observer that rotates uniformly about the $z$ axis on a circular orbit of constant radius $r$. The speed of the circular motion is $v = r \,\Omega$ and the observer's proper time is $\tau = t/\gamma$, where $\gamma$ is the corresponding Lorentz factor. As before, we assume that $\tau = 0$ when $t = 0$. The frequency content of incident radiation as measured by the rotating observer has been treated in detail in Ref.~\cite{BMB}.   If the wave is incident along the $z$ axis, we find via the Fourier analysis of the measured field that the frequency of the incident radiation in terms of the proper time of the observer is given by 
\begin{equation}\label{L36}
\omega' = \gamma (\omega \mp \, \Omega)\,, 
\end{equation}
where the Lorentz factor takes due account of time dilation. Equation~\eqref{L36} should be compared and contrasted in this case with the transverse Doppler effect given by $\omega'_{\rm D} = \gamma \omega$.

For the case of oblique incidence,
\begin{equation}\label{L37}
\omega' = \gamma (\omega -M \Omega)\,, 
\end{equation}
where $\hbar M$, $M = 0, \pm 1, \pm 2, ..., $ is the component of the total angular momentum of the electromagnetic radiation field along the rotation axis of the observer~\cite{BMB}. Let us multiply Eq.~\eqref{L37} by $\hbar$ and write the resulting equation in the WKB approximation as
\begin{equation}\label{L38}
E' = \gamma (E - \boldsymbol{\Omega} \cdot \mathbf{J})\,, 
\end{equation}
where 
\begin{equation}\label{L39}
\mathbf{J} = \mathbf{r} \times \mathbf{p}  + \mathbf{S}\, 
\end{equation}
 is the total angular momentum of the photon with momentum $\mathbf{p} = \hbar\, \mathbf{k}$ and spin $\mathbf{S}$. Thus, 
\begin{equation}\label{L40}
E' = \gamma (E - \mathbf{v} \cdot \mathbf{p}) - \gamma \mathbf{S}\cdot \boldsymbol{\Omega}\,, 
\end{equation}
where the energy due to spin-rotation coupling takes due account of time dilation. 

An important consequence of the helicity-rotation coupling is that an energy shift occurs when electromagnetic waves of definite helicity pass through a uniformly rotating half-wave plate. It appears that helicity-rotation coupling first came to light through the corresponding frequency shift of electromagnetic radiation in microwave experiments~\cite{Allen}. It was later studied in the optical regime by a number of investigators~\cite{G+A, Garetz, SKS, MNHS, 1Courtial, 2Courtial}. For further references and background material, see, for instance,~\cite{LCT, LZZ, ZFD, FPW, DLLL, CWL, E+E} and the references therein. The topic of energy shift as radiation passes through a spin flipper is briefly discussed in Appendix A. 

\section{Nonlocal Relativity}

The standard theory of relativity is based on the pointwise application of the Lorentz transformation for accelerated systems. This approach results, for wave phenomena, in the invariance of the phase of the wave under Lorentz transformations from which the standard Doppler formula is derived. On the other hand, if the pointwise measured field is subjected to the temporally nonlocal operation of Fourier analysis, one finds the more general results involving helicity-rotation coupling described above. In this case, phase invariance is violated. Furthermore, the photon helicity-rotation coupling is an example of the general coupling of intrinsic spin with rotation that is due to the \emph{inertia of intrinsic spin}~\cite{BM3}. In quantum mechanics, mass and spin determine the inertial properties of a particle, since they characterize the irreducible unitary representations of the Poincar\'e group~\cite{Wigner}. The inertial aspects of  mass are well known from Newtonian physics. The inertial aspects of intrinsic spin are independent of the inertial mass and depend solely on intrinsic spin. The inertial aspects of intrinsic spin have been the subject of recent investigations; see~\cite{Mashhoon:2015nea} and the references cited therein. There is much observational support for the general phenomenon of spin-rotation coupling; for instance, it is known as the ``phase wrap-up" in the Global Positioning System (GPS)~\cite{Ashby}. For recent developments in this general  subject,  see~\cite{DSH, DDKWLSH, Yu:2022vjn} and the references therein. 

There is, however, a fundamental difficulty here that must now be addressed. It follows from Eqs.~\eqref{L15} and~\eqref{L36} that an observer that rotates uniformly in the positive sense with angular speed $\omega$ about the direction of propagation of a plane monochromatic electromagnetic wave of frequency $\omega$ and positive helicity measures a frequency $\omega' = 0$; that is, the wave stands completely still for the class of all uniformly rotating observers with $r \ge 0$.  The measured electromagnetic field is oscillatory in space but independent of time. What happens in this case of rotational Doppler effect (RDE) is reminiscent of pre-relativistic Doppler effect where an observer moving at the speed of light along the direction of propagation could stay completely at rest with the wave. This circumstance was mentioned by Einstein in his autobiographical notes; see page 53 of~\cite{EinAN}. This problem does not occur in the relativistic Doppler formula given by Eq.~\eqref{I6} for inertial observers or Eq.~\eqref{J8} for accelerated observers; that is, the measured frequency will vanish only if the frequency of the incident wave vanishes. 

To overcome the problem in the present case, we must reconsider our assumption regarding the \emph{pointwise} measurement of the electromagnetic field. This is impossible according to Bohr and Rosenfeld, who have argued that only spacetime averages of electromagnetic fields can be measured.  

\subsection{The Bohr-Rosenfeld Principle}

Consider the presence of an electromagnetic field in a global inertial frame in Minkowski spacetime. In Cartesian coordinates $x^\mu = (ct, \mathbf{x})$, the electric and magnetic fields can be expressed as 
$\mathbf{E} (t, \mathbf{x})$ and  $\mathbf{B} (t, \mathbf{x})$, which are formally defined via the projection of the Faraday tensor $F_{\mu \nu}$ on the tetrad frame $e^{\mu}{}_{\hat \alpha}= \delta^{\mu}_{\hat \alpha}$ of the inertial observers at rest in the background frame.   

To simplify matters, let us imagine the actual measurement of the electric field by the inertial observers, while ignoring the presence of the magnetic field. A macroscopic body of linear size $\sim L$ and volume $V \sim L^3$ with constant uniform charge density $\rho$ is placed in the electric field. The body experiences a force given by the Lorentz force law
\begin{equation}\label{N1}
\frac{d \mathbf{P}}{dt} = \rho \int_V \mathbf{E} (t, \mathbf{x})\,d^3x\,, 
\end{equation}
where $\mathbf{P}$ is the momentum of the body. The inertial observers measure this momentum at the initial ($t'$) and final ($t''$) stages of the experiment that lasts for a total time $T = t'' - t'$. Hence, integrating Eq.~\eqref{N1} from $t'$ to $t''$, we get
\begin{equation}\label{N2}
 \mathbf{P}'' - \mathbf{P}' = \rho \int_{t'}^{t''} \int_V \mathbf{E} (t, \mathbf{x})\,d^3x = \rho <\mathbf{E}> TV\,. 
\end{equation}
What is measured is the spacetime average of the electric field given by
\begin{equation}\label{N3}
<\mathbf{E}> \, = \frac{1}{\Delta} \int_\Delta \mathbf{E} \,d^4x\,,
\end{equation}
where $\Delta = c\,TV$. This simple argument due to Bohr and Rosenfeld~\cite{B+R} assumes that momentum measurements by the inertial observers last for a time period $\ll T$ and cause displacements that are $\ll L$. 

\subsection{Nonlocality of Accelerated Observers}

The considerations of Bohr and Rosenfeld may appear somewhat innocuous for the measurements of inertial observers in Minkowski spacetime. Repeating the same arguments for accelerated observers, on the other hand, would present special difficulties. We therefore adopt a different indirect approach. The Bohr-Rosenfeld argument can be taken to mean that the measurement of the electromagnetic field by an accelerated observer requires the consideration  of the observer's past world line. That is, we must replace the pointwise measurement of the electromagnetic field by an accelerated observer with a nonlocal ansatz that involves a certain average of the field over the past world line of the accelerated observer. To this end, it is first necessary to discuss briefly the relativistic theory of accelerated systems. There are intrinsic spatial and temporal scales associated with accelerated observers; for instance, Earth-bound observers have translational and rotational acceleration lengths of $c^2/g_{\oplus} \approx 1$ light year and $c/\Omega_{\oplus} \approx 28$ astronomical units, respectively. 

\subsection{Acceleration Scales}

In Minkowski spacetime, we consider an arbitrary accelerated observer with proper time $\eta$ following a path  $x^\mu(\eta)$ given by inertial  Cartesian coordinates. The observer carries along its world line an adapted orthonormal tetrad frame $\xi^{\mu}{}_{\hat \alpha}(\eta)$ for measurement purposes, where $\xi^{\mu}{}_{\hat 0} = \mathfrak{u}^\mu = dx^\mu/d\eta$ is the timelike unit 4-velocity vector of the accelerated observer. The manner in which the observer carries its tetrad frame along its world line is specified via 
\begin{equation}\label{N4}
\frac{d \xi^{\mu}{}_{\hat \alpha}}{d\eta} = \Phi_{\hat \alpha}{}^{\hat \beta}\, \xi^{\mu}{}_{\hat \beta}\,.
\end{equation}
The orthonormality of the moving tetrad frame implies that the acceleration tensor $\Phi_{\hat \alpha \hat \beta}(\eta)$ is antisymmetric, namely, $\Phi_{\hat \alpha \hat \beta} = - \Phi_{\hat \beta \hat \alpha}$. Therefore, $\Phi_{\hat \alpha \hat \beta}$ is analogous to the Faraday tensor and can be represented in terms of its ``electric" and ``magnetic" components. The ``electric" part is given by
\begin{equation}\label{N5}
 \Phi_{\hat 0 \hat i} = \mathfrak{A}_{\hat i} = \mathfrak{A}_\mu \, \xi^{\mu}{}_{\hat i}\,,
\end{equation}
where $\mathfrak{A}^\mu = d\mathfrak{u}^\mu /d\eta$, $\mathfrak{A}_\mu \mathfrak{u}^\mu = 0$,  is the spacelike translational acceleration of the observer. Similarly, the magnetic part can be expressed as
\begin{equation}\label{N6}
 \Phi_{\hat i \hat j} = \epsilon_{\hat i \hat j \hat k}\, \Omega^{\hat k}\,, 
\end{equation}
where $\Omega^{\hat k}$ is the locally measured angular velocity of the rotation of the observer's local spatial frame with respect to a nonrotating (i.e. Fermi-Walker transported) frame. The measured components of the acceleration tensor $\Phi_{\hat \alpha \hat \beta}$ have dimensions of 1/length and can be employed to define the translational and rotational acceleration scales of the observer~\cite{BM2}. Using 
\begin{equation}\label{N7}
\Phi^{\mu \nu} =  \Phi^{\hat \alpha \hat \beta}\,\xi^{\mu}{}_{\hat \alpha}\,\xi^{\mu}{}_{\hat \beta}\,, \qquad  \Omega^\mu = \Omega^{\hat k}\, \xi^{\mu}{}_{\hat k}\,,
\end{equation}
where $\Omega^\mu$ is a spacelike 4-vector such that $\mathfrak{u}_\mu \,\Omega^\mu= 0$, we find
\begin{equation}\label{N8}
\Phi^{\mu \nu} =  \mathfrak{A}^{\mu} \mathfrak{u}^{\nu} -  \mathfrak{A}^{\nu} \mathfrak{u}^{\mu} + \epsilon^{\mu \nu \rho \sigma} \mathfrak{u}_\rho \Omega_\sigma\,.
\end{equation}
In our convention, the alternating tensor is in general such that $\epsilon_{\hat 0 \hat 1 \hat 2 \hat 3} = 1$. 

\subsection{Nonlocal Ansatz}

Regarding the arbitrary accelerated observer with proper time $\eta$ and orthonormal tetrad frame $\xi^{\mu}{}_{\hat \alpha}(\eta)$ under consideration in the previous subsection, let us assume that for $\eta \le \eta_0$ its motion is uniform and acceleration begins at $\eta =  \eta_0$. Thus, for $\eta >  \eta_0$, the observer along its world line passes through a continuous infinity of locally comoving inertial observers. The incident external field under consideration is assumed to be $\mathbb{A}_\mu$ and what an instantaneously comoving inertial observer measures along the world line is $\mathbb{A}_{\hat \alpha}(\eta)$ given by 
\begin{equation}\label{N9}
\mathbb{A}_{\hat \alpha}(\eta) = \mathbb{A}_{\mu}(\eta)\,\xi^{\mu}{}_{\hat \alpha}(\eta)\,. 
\end{equation}
On the other hand, what is actually measured by the accelerated observer is $\mathbb{B}_{\hat \alpha}(\eta)$. The most general relationship between $\mathbb{B}_{\hat \alpha}(\eta)$ and $\mathbb{A}_{\hat \alpha}(\eta)$ that is consistent with linearity and causality is 
\begin{equation}\label{N10}
\mathbb{B}_{\hat \alpha}(\eta) = \mathbb{A}_{\hat \alpha}(\eta) + \Theta(\eta - \eta_0) \int_{\eta_0}^{\eta} \mathbb{K}_{\hat \alpha}{}^{\hat \beta}(\eta')\,\mathbb{A}_{\hat \beta}(\eta')\,d\eta'\,,
\end{equation}
where $\Theta$ is the unit step function such that $\Theta(t) = 0$ for $t \le 0$ and $\Theta(t) = 1$ for $t > 0$. Ansatz~\eqref{N10} goes beyond the pointwise field measurement by including a certain average over the accelerated world line with a weight function $\mathbb{K}(\eta)$ that presumably depends upon the observer's acceleration. That is, when the acceleration is turned on at $\eta = \eta_0$, $\mathbb{B}_{\hat \alpha}(\eta_0) = \mathbb{A}_{\hat \alpha}(\eta_0)$; however, for $\eta > \eta_0$, there is in addition a certain cumulative memory of past acceleration involving a kernel $\mathbb{K}(\eta)$.  If the acceleration is turned off at some future time $\eta_f > \eta_0$, there will remain a constant memory of past acceleration for $\eta > \eta_f$ that is entirely benign. 

Along these lines, a nonlocal relativistic theory of accelerated observers was first proposed in 1993~\cite{Mashhoon:1993zz}. The main nonlocal ansatz is a Volterra integral equation and the unique correspondence between $\mathbb{B}_{\hat \alpha}(\eta)$ and $\mathbb{A}_{\mu}(\eta)$ in the space of functions of physical interest is guaranteed by the Volterra-Tricomi theorem~\cite{Mashhoon:1993zz, Mashhoon:2008vr}. 

To determine kernel $\mathbb{K}$, we assume that for an incident radiation field, the accelerated observer will always detect a time-dependent field; otherwise, if $\mathbb{B}_{\hat \alpha}$ turns out to be constant in time $\eta$, the incident field $\mathbb{A}_\mu$ must have been a constant field in the first place, just as in the case of the relativistic Doppler effect. That is, in this nonlocal theory no observer can ever stay at rest with respect to a pure radiation field. A constant $\mathbb{B}_{\hat \alpha}$ thus uniquely corresponds to a constant $\mathbb{A}_{\mu}$ in accordance with the Volterra-Tricomi theorem.  To implement this idea, let us therefore assume that $\mathbb{B}_{\hat \alpha}$ and $\mathbb{A}_\mu$ are constants for $\eta \ge \eta_0$.  Then, 
\begin{equation}\label{N10a}
\mathbb{B}_{\hat \alpha}(\eta) = \mathbb{B}_{\hat \alpha}(\eta_0) = \mathbb{A}_{\hat \alpha}(\eta_0)= \mathbb{A}_{\mu}(\eta_0)\,\xi^{\mu}{}_{\hat \alpha}(\eta_0)\,, 
\end{equation}
by our assumption and Eq.~\eqref{N10}. In this case, Eq.~\eqref{N10} can be simplified using Eq.~\eqref{N9} and for $\eta > \eta_0$ can be expressed for arbitrary constant $\mathbb{A}_\mu$ as
\begin{equation}\label{N11}
\xi^{\mu}{}_{\hat \alpha}(\eta_0) = \xi^{\mu}{}_{\hat \alpha}(\eta) +  \int_{\eta_0}^{\eta} \mathbb{K}_{\hat \alpha}{}^{\hat \beta}(\eta')\,\xi^{\mu}{}_{\hat \beta}(\eta')\,d\eta'\,.
\end{equation}
Taking the derivative of this relation with respect to $\eta$, we find
\begin{equation}\label{N12}
-\frac{d \xi^{\mu}{}_{\hat \alpha}(\eta)}{d\eta} = \mathbb{K}_{\hat \alpha}{}^{\hat \beta}(\eta)\, \xi^{\mu}{}_{\hat \beta}(\eta)\,.
\end{equation}
Comparing this relation with Eq.~\eqref{N4}, we conclude
\begin{equation}\label{N13}
\mathbb{K}_{\hat \alpha}{}^{\hat \beta}(\eta) = -\Phi_{\hat \alpha}{}^{\hat \beta}(\eta)\,;
\end{equation}
that is, the kernel of our nonlocal ansatz is minus the measured acceleration tensor of the observer. Further developments of these ideas are contained in~\cite{Mashhoon:2008vr, BMB, Mashhoon:2020mgw} and the references cited therein.  

In the rest of this paper, we work out the implications of this nonlocal theory for the observation of helicity-rotation coupling by the noninertial observer, introduced in Section III, that remains at rest at the origin of spatial coordinates in Minkowski spacetime. 

\section{Noninertial Observer: Nonlocal Ansatz}

We would like to find out how the nonlocal theory of accelerated systems developed in the previous section ameliorates the basic difficulty of the local theory. Therefore, we return to the problem of frequency measurements of the noninertial observer of Section III and assume that the observer is permanently at rest at the origin of spatial coordinates and for $t \le 0$ employs standard inertial Cartesian axes for measurement purposes.  However, for $t > 0$, the observer refers its measurements to the system that rotates uniformly about the $z$ axis and is given by Eqs.~\eqref{L1} and~\eqref{L2}. A plane electromagnetic wave of frequency $\omega$, definite helicity and 4-potential $A_\mu$ given by Eq.~\eqref{L3} is incident along the $z$ axis. Moreover, the noninertial observer's tetrad frame $e'^{\mu}{}_{\hat \alpha}$ is given by Eqs.~\eqref{L6}--\eqref{L9} for $t > 0$ and the electromagnetic field  locally measured by the comoving inertial observers is $A'_{\hat \alpha} = A_\mu \,e'^{\mu}{}_{\hat \alpha}$ as in Eq.~\eqref{L5}. The field actually measured by the noninertial observer is in this case $\mathcal{A}_{\hat \alpha} (t)$ given by our general nonlocal ansatz~\eqref{N10}, namely, 
\begin{equation}\label{N14}
\mathcal{A}_{\hat \alpha}(t) = A'_{\hat \alpha}(t) - \Theta(t) \int_{0}^{t} \Phi_{\hat \alpha}{}^{\hat \beta}(t')\,A'_{\hat \beta}(t')\,dt'\,,
\end{equation}
in accordance with the considerations of the previous section. 

The measured components of the acceleration tensor can be simply worked out in this case and we find that its nonzero components are given by
\begin{equation}\label{N15}
\Phi_{\hat 1 \hat 2} = - \Phi_{\hat 2 \hat 1} = \Omega\,.
\end{equation}
For the incident field~\eqref{L3} that propagates along the $z$ axis, $A'_{\hat \alpha}$ has been worked out in Section III and is given by  $A'_{\hat 0} = A'_{\hat 3} = 0$ and
\begin{equation}\label{N16}
A'_{\hat 1}(t) = a_{\pm} \,e^{-i(\omega \mp \Omega)t}\,, \qquad A'_{\hat 2} =  \pm i\,A'_{\hat 1}\,.
\end{equation}
It is then straightforward to employ Eq.~\eqref{N14} to show that $\mathcal{A}_{\hat 0} = \mathcal{A}_{\hat 3} = 0$ and
\begin{equation}\label{N17}
\mathcal{A}_{\hat 1}(t) = \frac{a_{\pm}}{1\mp \tfrac{\Omega}{\omega}} \,\left[e^{-i(\omega \mp \Omega)t} \mp \,\tfrac{\Omega}{\omega}\right]\,, \qquad \mathcal{A}_{\hat 2} = \pm i\, \mathcal{A}_{\hat 1}\,.
\end{equation}
The measured frequency is still $\omega' = \omega \mp \Omega$ as in Eq.~\eqref{L15}; however, the measured amplitude in the nonlocal theory picks up a helicity-dependent factor
\begin{equation}\label{N18}
\frac{\omega}{\omega'_{\pm 1}} = \frac{1}{1\mp \tfrac{\Omega}{\omega}}\,.
\end{equation}
In practice,  $\Omega \ll \omega$, and the measured amplitude slightly increases (decreases) if the observer rotates in the positive sense about the direction of propagation of positive (negative) helicity radiation. For incident radio waves of frequency $1$ GHz, we have $\Omega / \omega = 10^{-7}$ for an observer that rotates one hundred times per second. It is hoped that future observations can verify this consequence of helicity-rotation coupling. In particular, it would be interesting if the nonlocal helicity-dependent  amplitude effect could be detected within the context of RDE. 

The nonlocal theory was introduced to ensure that electromagnetic radiation could never stand completely still with respect to an observer. We recall that $\Omega = \omega$ caused the problem in the case of incident positive-helicity radiation; on the other hand, in the nonlocal theory the wave amplitude is no longer constant as $\Omega \to \omega$, since
\begin{equation}\label{N19}
\lim_{\Omega \to \omega} \frac{e^{-i(\omega - \Omega)t} - \tfrac{\Omega}{\omega}}{1 - \tfrac{\Omega}{\omega}} = 1 - i\, \omega t\,.
\end{equation}
Hence, in this resonance case the nonzero components of $\mathcal{A}_{\hat \alpha}(t)$ are given by
\begin{equation}\label{N20}
\mathcal{A}_{\hat 2}(t) = i\,\mathcal{A}_{\hat 1}(t) = a_{+} (1 - i\, \omega t)\,.
\end{equation}
In this way, the nonlocal approach avoids the fundamental difficulty of the local theory.  

In the limit of large quantum numbers, there is a correspondence between the behavior of electrons in atoms and accelerated observers of classical physics. Indeed, using Bohr's correspondence principle, it is possible to show that the consequences of nonlocal theory given by Eqs.~\eqref{N18} and~\eqref{N20} are consistent with the quantum theory~\cite{BM4}. 

\subsection{Oblique Incidence: Nonlocal Ansatz}

In the case of oblique incidence, we use Eq.~\eqref{N14} for $t > 0$ with $A'_{\hat \alpha}$ given by Eqs.~\eqref{L22a}--\eqref{L24a} to get
\begin{equation}\label{N21}
\mathcal{A}_{\hat 1} = \frac{\omega}{\omega'_{+1}}\,\psi_{(\hat 1, +1)}\, \left[e^{-i\,\omega'_{+1} t} - \tfrac{\Omega}{\omega}\right]  + \frac{\omega}{\omega'_{-1}}\,\psi_{(\hat 1, -1)}\, \left[e^{-i\,\omega'_{-1} t} - \tfrac{\Omega}{\omega}\right]\,,
\end{equation}
\begin{equation}\label{N22}
\mathcal{A}_{\hat 2} = \frac{\omega}{\omega'_{+1}}\,\psi_{(\hat 2, +1)}\, \left[e^{-i\,\omega'_{+1} t} - \tfrac{\Omega}{\omega}\right]  + \frac{\omega}{\omega'_{-1}}\,\psi_{(\hat 2, -1)}\, \left[e^{-i\,\omega'_{-1} t} - \tfrac{\Omega}{\omega}\right]\,,
\end{equation}
and
\begin{equation}\label{N23}
\mathcal{A}_{\hat 3} = \psi_{(\hat 3, 0)}\, e^{-i\,\omega t}\,,
\end{equation}
where $\omega'_{\pm1} = \omega \mp \Omega$. The observer measures the same frequency spectrum as before; however, the amplitudes can change in accordance with   
\begin{equation}\label{N24}
\psi_{(\hat i, m')} \to  \frac{\omega}{\omega'_{m'}}\, \psi_{(\hat i, m')}\,.
\end{equation}
Therefore, the only amplitude change that occurs involves $\omega'_{\pm1}$, in agreement with Eq.~\eqref{N18}. 

In the nonlocal theory, the average frequency measured by the noninertial observer may be calculated as in Eq.~\eqref{L29} taking due account of the variation in amplitude given by Eq.~\eqref{N24}; that is, 
\begin{equation}\label{N25}
< \omega' > \, = \frac{\sum |\psi_{(\hat i, m')}|^2/\omega'_{m'}}{\sum |\psi_{(\hat i, m')}|^2/\omega'^2_{m'}}\,. 
\end{equation}
It follows that 
\begin{equation}\label{N26}
\frac{< \omega' >}{\omega}  \, = \frac{\mathbb{N}}{\mathbb{D}}\,, 
\end{equation}
where
\begin{equation}\label{N27}
\mathbb{N} := 1 \pm \epsilon \cos \theta - \epsilon^2 ( 1 + \tfrac{1}{2} \sin^2 \theta) \mp \epsilon^3 \cos \theta + \tfrac{1}{2} \epsilon ^4 \sin^2 \theta\,,
\end{equation}
\begin{equation}\label{N28}
\mathbb{D} := 1 \pm\,2 \epsilon \cos \theta + \epsilon^2 ( 1 - \tfrac{3}{2} \sin^2 \theta) + \tfrac{1}{2} \epsilon ^4 \sin^2 \theta\,, 
\end{equation}
 \begin{equation}\label{N29}
\epsilon := \frac{\Omega}{\omega}\,.
\end{equation}
In powers of $\epsilon \ll 1$, we find
\begin{equation}\label{N30}
\frac{< \omega' >}{\omega}  \, = 1 \mp \epsilon \cos \theta  - \epsilon^2 \sin^2 \theta \pm \frac{1}{2} \epsilon^3 \sin^2 \theta \cos \theta + O(\epsilon^4)\,,
\end{equation}
so that we recover the simple helicity-rotation coupling to first order in $\epsilon$. 

We have chosen the observer to be fixed at the origin of spatial coordinates, i.e. $r = 0$. In this way, the photon orbital angular momentum vanishes and we only deal with the photon helicity.  For a nonlocal treatment involving rotating observers with $r > 0$, see~\cite{BM5}.

\section{Helicity-Dependent Amplitude Effect}

It turns out that the result of the nonlocal ansatz expressed by Eq.~\eqref{N18} is quite general and may be amenable to experimental research. If the observer rotates with angular velocity $\Omega$ in the positive sense about the direction of propagation of  incident positive (negative) helicity plane electromagnetic wave of  frequency $\omega \gg \Omega$, the amplitude of the wave as measured by the rotating observer is enhanced (diminished) by $(1\mp \Omega/\omega)^{-1} \approx 1 \pm  \Omega/\omega$. That is, the amplitude is slightly larger (smaller) if the electromagnetic field rotates in the same (opposite) sense as the observer. We explore this \emph{nonlocal} effect for an incident linearly polarized wave in this section. 

To see explicitly the significance of this nonlocal helicity-dependent amplitude effect, we consider a linearly polarized plane electromagnetic wave incident along the $z$ axis with vector potential $A_\mu = (0, \mathbf{A})$, where 
\begin{equation}\label{M1}
\mathbf{A} = \tilde{a}\,\hat{\mathbf{x}}\, e^{-i\,\omega (t - z/c)}\,.
\end{equation}
Here $\tilde{a}$ is a constant complex amplitude. Assuming the locality postulate, the noninertial observer fixed at the origin of spatial coordinates with adapted tetrad frame~\eqref{L5}--\eqref{L9} would measure a vector potential given by  $A'_{\hat 0} = 0$, $A'_{\hat 3} = 0$ and 
\begin{equation}\label{M2}
 A'_{\hat 1} = \tilde{a}\, e^{-i\,\omega t}\, \cos\Omega t\,, \qquad A'_{\hat 2} = - \tilde{a}\, e^{-i\,\omega t}\, \sin\Omega t\,.
\end{equation}
Let us note that 
\begin{equation}\label{M3}
 A'_{\hat 1}\, \hat{\mathbf{x}}' + A'_{\hat 2}\, \hat{\mathbf{y}}' = \tilde{a}\, \hat{\mathbf{x}} \,e^{-i\,\omega t}\,,
\end{equation}
where we have used Eq.~\eqref{L1}. Assuming that the electromagnetic field is pointwise measured by the noninertial observer, the plane of linear polarization of the incident wave would appear to rotate in the negative sense with angular speed $\Omega$ about the $z$ axis. 

On the other hand, employing the nonlocal ansatz~\eqref{N14}, we find instead of Eq.~\eqref{M2},  
\begin{equation}\label{M4}
 \mathcal{A}_{\hat 1} = \frac{\tilde{a}}{1-\epsilon^2} \left[(\cos\Omega t + i\,\epsilon\, \sin\Omega t) e^{-i\,\omega t} - \epsilon^2 \right]\,, 
\end{equation}
\begin{equation}\label{M5}
 \mathcal{A}_{\hat 2} = \frac{\tilde{a}}{1-\epsilon^2} \left[(- \sin\Omega t + i\,\epsilon\, \cos\Omega t) e^{-i\,\omega t} - i\,\epsilon \right]\,,
\end{equation}
where $\epsilon =\Omega/\omega$, as before. In this case, 
\begin{equation}\label{M6}
 \mathcal{A}_{\hat 1}\, \hat{\mathbf{x}}' + \mathcal{A}_{\hat 2}\, \hat{\mathbf{y}}' = \frac{\tilde{a}}{1-\epsilon^2} \left[(\hat{\mathbf{x}} + i\,\epsilon\, \hat{\mathbf{y}}) e^{-i\,\omega t} - \epsilon^2\,\hat{\mathbf{x}}' - i\,\epsilon\, \hat{\mathbf{y}}' \right]\,,
\end{equation}
which should be compared and contrasted with Eq.~\eqref{M3}. In practice, $\epsilon \ll 1$ and to first order in $\epsilon$,  
\begin{equation}\label{M7}
\tilde{a}\, (\hat{\mathbf{x}} + i\,\epsilon\, \hat{\mathbf{y}})\, e^{-i\,\omega t}\,
\end{equation}
in Eq.~\eqref{M6} represents an elliptically polarized wave with ellipticity $\epsilon \ll 1$. Therefore, an incident linearly polarized electromagnetic wave appears to the noninertial observer to be slightly elliptically polarized while precessing in the negative sense about the direction of incidence $z$ with angular speed $\Omega$. To verify such effects observationally, it is clearly necessary to measure the ellipticity of electromagnetic radiation to rather high accuracy. In this connection, see, for instance, Ref.~\cite{KDL} and references therein.

\section{Discussion}

The photon helicity-rotation coupling discussed in this paper is an aspect of the general phenomenon of the coupling of intrinsic spin with rotation as well as the gravitomagnetic fields of rotating systems~\cite{Mashhoon:2023idh}. 

The nonlocal approach we employ here has a marked resemblance to the nonlocal electrodynamics of media~\cite{Jackson, LD+EM, HeOb}. To introduce history dependence into the theories of accelerated systems and gravitational fields, it seems prudent to follow the method that has been successful in the physics of continuous media. The state of a physical system can be history dependent; for instance, magnetic materials exhibit the phenomenon of hysteresis. This general subject has a long history; see, for instance,~\cite{Poi, Hop} and the references therein. For the nonlocal extension of Einstein's theory of gravitation along these lines, see~\cite{Hehl:2008eu, Hehl:2009es, BMB,  VanDenHoogen:2017nyy}.

It seems challenging to find a physical circumstance that could allow the measurement of nonlocal aspects of accelerated systems~\cite{BaMa, Bremm:2015wmv}.  In this connection, the observational implications of the nonlocal helicity-dependent amplitude effect discussed in the previous section could be interesting. 

\appendix

\section{Energy Shifts via Spin-Rotation Coupling}

Imagine a  half-wave plate in the $(x, y)$ plane that rotates slowly and uniformly with angular speed $\Omega$ about the $z$ axis. An incident plane electromagnetic wave of frequency $\omega_{\rm in}$ and definite helicity propagates along the $z$ axis. The rotation of the half-wave plate is uniform; therefore, the spacetime within the rotating half-wave plate is stationary.  If we neglect time dilation, all observers at rest with respect to the half-wave plate measure a frequency $\omega' \approx \omega_{\rm in} \mp \, \Omega$ regardless of their distance from the $z$ axis in accordance with Eq.~\eqref{L36}; here, as before, the upper (lower) sign refers to incident positive (negative) helicity radiation. This remarkable fact implies that as the wave passes through the half-wave plate, its frequency remains constant and equal to $\omega'$ as measured by observers at rest in the half-wave plate but its helicity changes as it leaves the plate. Thus, observers at rest at the bottom of the half-wave plate measure a frequency 
\begin{equation}\label{A1}
\omega' \approx \omega_{\rm in}  \mp \Omega\, 
\end{equation}
for the \emph{incoming} wave, while observers at rest at the top of the half-wave plate measure the same frequency
\begin{equation}\label{A2}
\omega' \approx \omega_{\rm out}  \pm \Omega\, 
\end{equation}
for the \emph{outgoing} wave. It follows that 
\begin{equation}\label{A3}
\omega_{\rm out} - \omega_{\rm in} \approx \mp \,2 \,\Omega\,. 
\end{equation}
Multiplying this relation by $\hbar$, we get the general result~\cite{MNHS}
\begin{equation}\label{A4}
E_{\rm out} - E_{\rm in} \approx - 2\, \mathbf{S} \cdot \,\boldsymbol{\Omega}\, 
\end{equation}
for radiation passing through a spin-flipper. 

In the background inertial frame, the kinetic energy of the rotating half-wave plate in the nonrelativistic approximation is given by   
\begin{equation}\label{A5}
\mathcal{E} = \frac{1}{2} \mathcal{I}\, \Omega^2 = \frac{1}{2} \frac{\mathcal{J}^2}{\mathcal{I}}\,, 
\end{equation}
where $\mathcal{I}$ is the moment of inertia of the plate and $\mathcal{J} = \mathcal{I}\,\Omega$ is its angular momentum. Thus, assuming that the moment of inertia of the plate remains constant as the wave passes through, 
\begin{equation}\label{A6}
\frac{\partial\mathcal{E}}{\partial \mathcal{J}} =  \Omega\,. 
\end{equation}
For each photon of definite helicity that passes through the half-wave plate, the angular momentum of the plate changes by $\delta \mathcal{J} = \pm\,2 \hbar$, where the upper (lower) sign indicates incident positive (negative) helicity radiation; therefore, the corresponding change in the kinetic energy of the plate is given by $\delta \mathcal{E} = \pm\,2 \hbar \Omega$. Conservation of energy then requires that the change in the frequency of the photon be given by $\delta \omega = \omega_{\rm out} - \omega_{\rm in} = \mp \,2 \,\Omega$. It follows that the phenomenon of energy shift through helicity-rotation coupling is consistent with the laws of conservation of energy and angular momentum. 

For a corresponding treatment of the nonlocal helicity-dependent amplitude effect in the process of energy shift due to helicity-rotation coupling, see~\cite{BM6}.


\end{document}